# Enwrapped Perylene Bisimide Enables Room Temperature Polariton Lasing and Photonic Lattices

*Dominik Horneber†, Johannes Düreth†, Tim Schembri, Simon Betzold, Matthias Stolte, Sven Höfling, Frank Würthner*, and Sebastian Klembt**

D. Horneber, J. Düreth, Dr. S. Betzold, Prof. Dr. S. Höfling, Prof. Dr. S. Klembt
Technische Physik, Wilhelm-Conrad-Röntgen-Research Center for Complex Material Systems, and Würzburg-Dresden Cluster of Excellence ct.qmat, University of Würzburg, 97074 Würzburg, Germany
†These authors have contributed equally.
*E-mail: sebastian.klembt@uni-wuerzburg.de

T. Schembri, Dr. M. Stolte, Prof. Dr. F. Würthner
Institut für Organische Chemie and Center for Nanosystems Chemistry (CNC), University of Würzburg, 97074 Würzburg, Germany
*E-mail: frank.wuerthner@uni-wuerzburg.de

**Perylene bisimides (PBIs) are organic dyes with photoluminescence quantum yields (PLQY) close to unity in solution and great thermal and photo-chemical stability. These features alongside the tunability of their solid-state packing arrangement via chemical functionalization make this material class an excellent candidate for exciton-polariton lasing at room temperature. Polariton lasing is well understood in III-V semiconductors at cryogenic temperatures, however, the search for emitter materials for robust and versatile room temperature applications is ongoing. While e.g. perovskites and several organic materials have been identified to support polariton lasing, many of these materials lack tunability and long-term stability under ambient conditions. Here, we fabricate optical microcavities using a strongly enwrapped PBI chromophore with prevailing monomer-like absorption and emission properties in the solid state. Voluminous bay-**

**substituents prevent π-π-stacking induced PLQY-quenching, thereby enabling polariton lasing at room temperature. Additionally, photonic confinement in single hemispheric resonators is demonstrated leading to localized polaritonic modes with discrete energies, as well as optical lattices revealing distinct polaritonic band-structures. Due to the possibility of tunable properties by the precise control of the solid-state packing arrangement of PBI emitters, our results pave the way for polarization-dependent light-matter coupling, including topological photonic effects within oriented crystalline thin-film microcavity structures.**

## 1. Introduction

Microcavity exciton-polaritons (polaritons) are quasi-particles that originate from the strong coupling between excitons and photons in an optical microcavity [1-2]. Their photonic component gives polaritons a much-reduced effective mass compared to the electron ($\sim 10^{-5}\,m_e$) and a straightforward experimental access via spectroscopic methods, while their matter component leads to optical nonlinearities as well as controllability via magnetic fields and electrical charge implementation. The bosonic nature allows the transition to a nonequilibrium condensate with the macroscopic occupation of a single state, which leads to the emission of coherent light bypassing the necessity for population inversion [3-4]. Early implementations largely relied on conventional III-V and II-VI semiconductor microcavities [5-6], including electrical operation [7-9]. In recent years, exciting phenomena such as Bose-Einstein condensates [10], superfluidity [11] or the development of potential landscapes for the implementation of quantum simulators [12], transistors [13] and topological Chern insulators [14] have emerged as further fields of application for polariton condensates. Traditionally, the primary emissive material platform for such studies has been high-quality GaAs-based crystalline semiconductor microcavities that require state-of-the-art clean room facilities for their fabrication and operation is typically constrained to cryogenic temperatures. To make the interesting effects of polariton condensates practicable for applications, polariton condensation was also demonstrated at room temperature [15] in materials with a large exciton binding energy e.g. zinc oxides [16], perovskites [17] and several organics [18]. While strong coupling between excitons and microcavity photons using an organic emitter material was first shown with a polystyrene-doped Zn-porphyrin (4TBPPZn) film [19] and polariton lasing was first demonstrated in anthracene single-crystals [20], the field has broadened considerably and now includes a plethora of organic materials [21-22]. Prominent examples are small molecules [23], conjugated polymers [24], fluorescent proteins [25], 2D materials [26] and carbon nanotubes [27]. So far, organic microcavity polaritons have also been used to realize devices such as e.g. transistors [28], few photon detectors [29] and narrow-band organic light-emitting diodes (OLEDs) [30]. While a wide range of materials has been investigated, the search for an ambient condition long-term stable, robust, versatile and tunable emitter material is still ongoing.

In the following work, we investigate strong light-matter interaction in an optical microcavity containing a perylene bisimide (PBI) solid-state emitter. This class of organic dyes is well-known for their high photostability and photoluminescence quantum yield (PLQY) close to unity in solution [31]. Due to their high electron affinity as well as outstanding charge transport properties as n-type semiconductors, PBIs also find many applications in transistors [32-35], OLEDs [36-38], lasers [39-40] or photovoltaics [41-45]. However, whilst changing molecular optical properties via chemical functionalization at different positions at the chromophore π-core is possible for this material class, the quenching of their fluorescence in the solid state so far limited many desired applications. To prevent undesired PLQY quenching in the solid state, effective shielding of the chromophore π-core in 1,7-disubstituted PBI derivatives [46-47] against molecular interactions proved successful. This enwrapment of the PBI enables almost monomer-like absorption and emission properties in the (crystalline) solid state. Over the last years, perylene bisimides [48-49] have been found to be a promising platform for studying light-matter interaction in optical microcavities at room temperature. So far, several works [50-56] using PBIs for strong light-matter interaction in different cavity geometries such as open cavities or solution-processed microcavities are reported. Although their tremendous potential for polariton lasing has been stated, an unequivocal proof of polariton lasing in the material class of PBIs is still elusive.

In this work, we unambiguously demonstrate strong light-matter coupling and exciton-polariton lasing in a planar optical microcavity consisting of dielectric distributed Bragg reflectors (DBRs) and a neat polycrystalline **PBI-1** layer. Furthermore, we implement different photonic potentials via lateral patterning of the cavity and exhibit discrete polaritonic modes as well as the formation of a polaritonic band structure. Our results show the great photonic flexibility of **PBI-1** based polaritonic devices and demonstrate the benefits of manipulating the excitonic and the photonic counterpart of light-matter quasiparticles independently.

## 2. Results

In this work, we use the solid-state emitter **PBI-1** [47] with two bulky 2,4,6-tris-(3,5-di-*tert*-butylphenyl)-phenoxy bay substituents, which successfully suppress molecular interactions e.g. exciton coupling in its crystalline solid state by increasing the center-to-center distance to about 15.73 Å. The enwrapment from the bay-positions as well as the central chromophore core of **PBI-1** highlighted in orange is depicted in Figure 1a. By these means quenching of the PLQY due to electronic interactions and the formation of non-radiative deactivation pathways is avoided. Accordingly, solid state PLQY up to 60% has been demonstrated in thin films as well as single-crystals with monomer-like absorption and emission properties in the visible spectrum [47]. A detailed scheme of the type of devices we fabricated is shown in **Figure 1b**. It consists of two distributed Bragg reflectors (DBRs), each of them made from nine $SiO_2/TiO_2$ pairs that were deposited on a quartz glass substrate by dual ion beam sputtering. The individual layer thicknesses are chosen in a way that the PBI's $S_0$-$S_1$ transition at about 561 nm is spectrally located in the center of the DBR stopband, and the sample can be pumped non-resonantly through the first high-energy Bragg minimum at 482 nm into the higher $A_{02}$ vibronic progression. To place the emitter layer in an anti-node of the cavity electric field, a $SiO_2$ spacer layer is deposited onto the bottom mirror during the sputtering process. The resulting distribution of the electric field in the microcavity structure as obtained by transfer-matrix method (TMM) calculations can be found in the supplementary **Figure S1**. The **PBI-1** layer with tunable thickness was spin-coated from 4 mg mL$^{-1}$ dichloromethane solution onto the bottom DBR with 1500 rpm for 30 sec and annealed at 150°C for 5 min to remove residual solvent as well as to improve crystallinity of the isotropic layer. Transmission electron microscope (TEM) and selected area electron diffraction (SAED) measurements for the characterization of the thin film can be found in **Figure S2** of the Supporting Information. The resulting molecular packing is depicted schematically in the inset of **Figure 1b**. The microcavity sample is completed by drop-casting 5μL of BSA in water (concentration 200mg/ml) as adhesion agent directly onto the PBI layer and placing the top DBR on top of it by applying mechanical pressure for 60s. Afterwards, the sample is stored under constant mechanical pressure of ~0.25 N cm$^{-2}$ for 15 h. This method provides a remaining thickness gradient of typically ~3 λ over the 1×1 mm$^2$ sample area and therefore a detuning gradient of the cavity mode resulting in tunable excitonic and photonic fractions of the polaritons. To find a suitable thickness of the solid-state emitter layer, several samples with **PBI-1** thicknesses between 10 nm and 60 nm have been produced and studied. While for low thicknesses the interaction between cavity photons and emitter is found to be in the weak coupling regime, for large thicknesses the self-absorption of the PBI´s $A_{00}$ transition limited the PL intensity of the sample. A trade-off between these two extrema is found to be around L = (25 ± 3) nm of PBI thickness. For each layer thickness photoluminescence (PL) and absorption measurements of the neat emitter material on a glass substrate are recorded for a precise determination of the spectral positions of the excitonic transitions. The corresponding spectra of a **PBI-1** layer with 25 nm thickness can be found in **Figure S3** of the Supporting Information and show the monomer-like peaks in absorption and PL. For angle-resolved reflection measurements of our **PBI-1** containing microcavities, the sample is illuminated with a white-light LED while for PL measurements the sample was excited with a tunable, pulsed laser source at 482 nm. In both cases, the incident light is focused onto the sample with a microscope objective with numerical aperture of NA=0.65 leading to a spot size of approximately 3 µm. The signal is spectrally resolved with a monochromator and displayed on a CCD camera. By placing lenses with focal length *f* in the detection path in 3*f* and for 4*f* configuration, one can change between *real-* and *reciprocal-*space imaging, respectively.

For a planar microcavity sample with a **PBI-1** thickness of 25 nm, large areas of the sample exhibited a cavity length of ~ λ/2 and resulted in one photonic mode over the entire stop-band width. Furthermore, strong coupling of the photonic and excitonic mode could be found. This was verified via angle-resolved white-light reflection measurements as shown in **Figure 1c** on the left side with the clear formation of and upper and lower polariton mode around the exciton energy. This agrees with TMM calculations using the measured dielectric constants of the neat 25 nm **PBI-1** emitter layer. The result is depicted in the middle part of **Figure 1c** and overlaps with the precise positions of both polaritonic modes extracted by Lorentzian fitting of the reflection data (black crosses). From the TMM simulation the thickness of the BSA layer in this area was determined to be *d*~57 nm. Fitting of the coupled harmonic oscillator (CHO) model to the data yields a vacuum Rabi splitting of $\hbar\Omega_R$ = 98 meV (see **Figure S3**). While in reflection the upper polariton branch is visible for high angle-values, in PL only the lower polariton branch is populated as shown in **Figure S4** together with the fitting results of the CHO model. The structure yields an experimental microcavity quality factor Q~1100 which is determined by the full-width half maximum (FWHM) of the most photonic lower polariton mode that could be measured in PL.

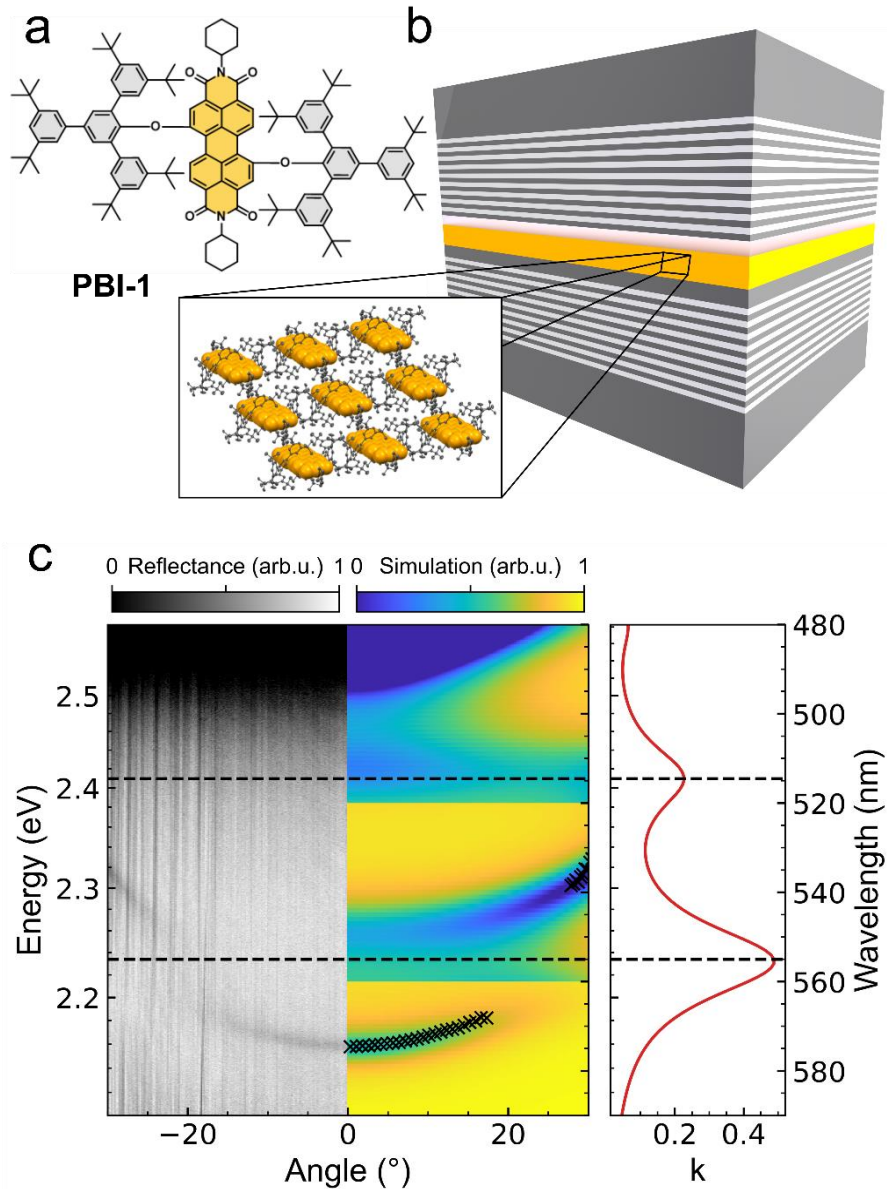

Figure 1. a) Chemical structure of the **PBI-1** emitter material. b) Scheme of a planar microcavity sample. It consists of nine SiO$_2$/TiO$_2$ pairs for the top and bottom DBR mirrors, a SiO2 spacer layer, a 25 nm thick **PBI-1** layer and a BSA layer acting as adhesion glue. The inset shows the molecular packing inside the polycrystalline emitter thin film. c) Angle-resolved reflection measurements and calculations of the microcavity sample exhibiting the formation of an upper and lower polariton branch around the excitonic transition that is nicely represented in measurements of the absorption coefficient k (right side).

For investigating the polariton lasing properties of our strongly coupled microcavities, we excite the microcavity sample with a laser spot of ~10 µm in diameter and systematically increase the pump power at 482 nm. At a power of 125 nJ/pulse, a dispersionless mode with high intensity shifts out of the polariton mode towards higher energies as shown in **Figure 2 a-c**. At this threshold power a nonlinear enhancement in the emission intensity as well as a strong decrease in the linewidth of the mode is observed which indicates the build-up of phase coherence (compare **Figure 2d**). Furthermore, a continuous blueshift of the mode to a maximum of 15 meV is visible around the threshold power. Intensity profiles of the lower polariton mode at several pumping power steps below and above threshold can be found in supplementary **Figure S5**. These features are hallmark signs for polariton lasing. It should be noted that while the blueshift in Wannier-Mott exciton systems is a result of Coulomb interaction, in organic emitters it is typically dominated by phase space filling resulting in a reduction of the Rabi splitting as discussed by several works [57-58]. For increasing optical pump power, multiple modes at lower energies become visible that are believed to be a consequence of disorder in packing structure and volume of the thin film.

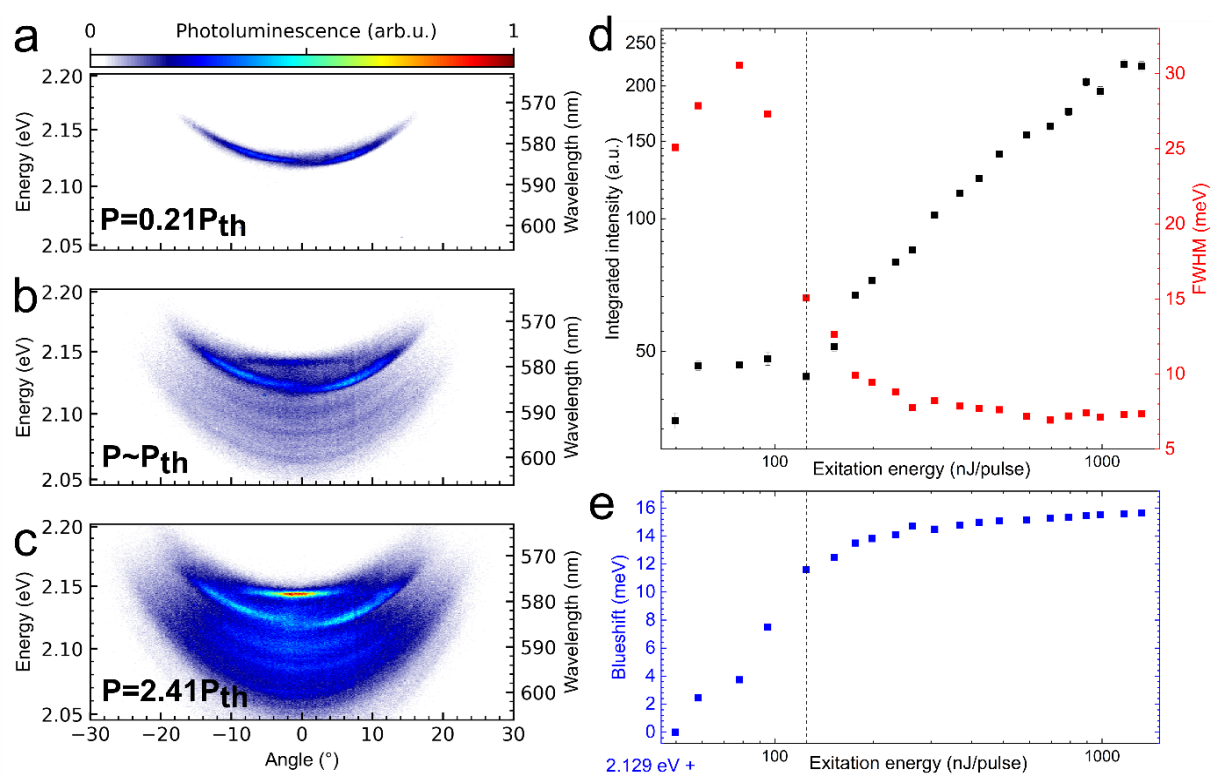

Figure 2. a)-c) Angle-resolved photoluminescence of a PBI-based microcavity below (a), at (b) and above (c) the threshold power for polariton lasing. When crossing the threshold, a dispersionless mode shifts in higher energy direction out of the polariton mode. d) Linewidth and input-output characteristics of the PBI lower polariton mode. At the threshold power at around 125 nJ/pulse, a strong decrease of the linewidth (red symbols) as well as a non-linear increase in intensity (black symbols) is observed. e) The blueshift of the lasing mode shows

the typical behavior of different slopes below and above the threshold power. Overall, a blueshift of 15 meV can be observed.

Over the last decade polariton lattices have grown in interest due to the possibility of emulating atomic lattices with exciton-polaritons at a microscopic level [59]. These artificial lattices can be built in virtually any 2D geometry, no matter if the lattice exists as a crystallographic lattice in nature or not. Therefore, one can simulate new lattice structures and study e.g. topological effects with the platform of microcavity polaritons. While for traditional III-V semiconductor materials complex etching techniques for creating e.g. honeycomb [60] or Lieb lattices [61] are required, for room-temperature system often the manipulation of the top dielectric DBR via focused ion beam milling is used to create the photonic confinement of the same lattices [62-63]. In our **PBI-1** based microcavities the bottom part with the 25 nm solid-state emitter layer stays equal to the other samples, and focused ion beam milling (FEI Helios NanoLab DualBeam) is used to nanofabricate dimples of hemi-elliptical indentations into a borosilicat substrate before coating it with the mirror pairs. The cavity was again completed by using BSA for glueing the structured top mirror onto the bottom DBR with the emitter layer.

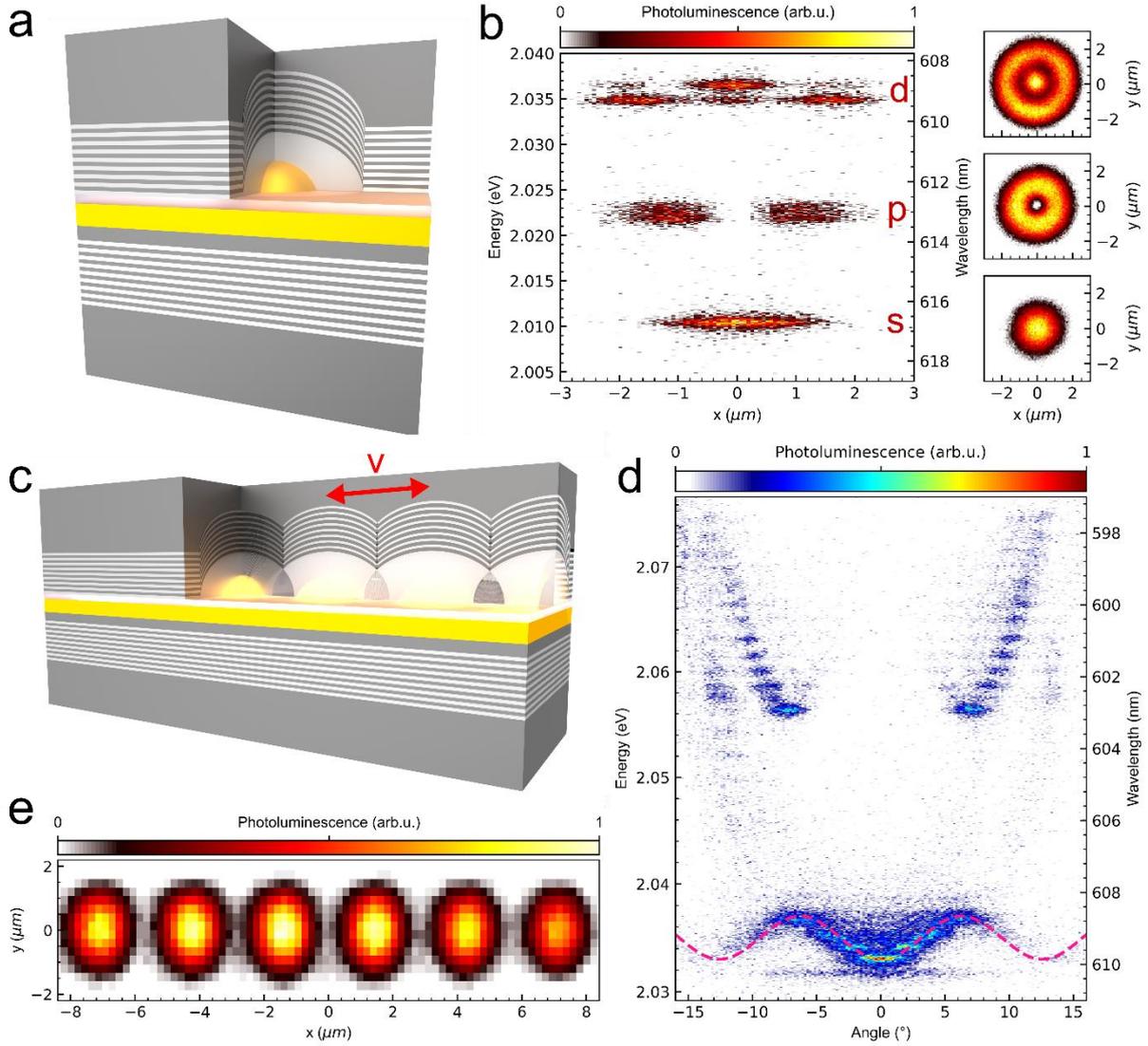

Figure 3. a) Schematic of the single photonic trap in a **PBI-1** based microcavity. b) Discrete dispersionless polariton modes in real-space caused by the photonic confinement. The inset shows the 2D real space image of each of the modes c) Drawing of hemispheres coupled to a linear chain with a coupling strength v, defined by the overlap of neighboring lenses. d) Angle-resolved band structure in PL of the linear chain with a band gap rising between binding and ant-binding modes. With dotted lines a fit of a tight-binding model for the s-band is marked. e) Real-space mode tomography of the binding s-mode of the linear chain showing the PL of each single trap.

PL measurements below the threshold power were performed on single photonic traps as schematically shown in **Figure 3a.** This type of hemispherical confinement is a well-known method for e.g. producing coherent single photons [64], investigating tunable polariton lasing [65] or polariton lattices based on $WS_2$ monolayers [66]. In this work, traps with depths of $t$~200 nm, diameters between $d$=3 µm and $d$=6 µm, and therefore radii of curvature (ROC=$d^2/4t$) up to 45 µm were investigated. **Figure 3b** shows the real-space mode profile of the first three out of five energy states that could be observed in a trap with diameter of 6 µm. The dispersionless modes have equidistant energies as well as an energy-splitting of the higher-order modes due to a splitting in transverse electric and transverse magnetic fields (TE/TM splitting). The Q-factor of the lowest energy state was experimentally obtained to be in the order of Q~3200. A similar mode spectrum has been observed by Dusel *et al.* [48] using a sophisticated but technologically challenging imprinting technique with a liquid crystalline PBI derivate.

To investigate the real-space shape of each mode, hyperspectral imaging was conducted. The real-space images of the first three energy-levels are depicted in the inset of **Figure 3b**. Since the hemispherical resonator is a photonic equivalent of a single atom, the confined Laguerre-Gaussian modes are shaped like the equivalent atomic orbitals. For the ground-state, referred to as s-state, the mode is radially symmetric and for the first excited state, called p-state, the mode shows the characteristic doughnut-like shape.

The next step towards artificial polariton lattices with a **PBI-1** based exciton-polariton system is the transition from zero-dimensional confinement in a hemispherical lens to one-dimensional confinement in a linear chain. This chain consists of a periodic arrangement of equidistant overlapping hemispherical traps as shown in Figure 3c. The coupling strength between the lattice sites is defined by the overlap of the hemispheres which is given as $v=a/d=0.6$ where *a* is the center-to-center distance and *d* is the diameter of a single lens. The sample was excited at a wavelength of 482 nm and a power below threshold as before, but with a line-shaped laser spot. The angle-resolved dispersion of a chain with $d$=4 µm diameter of the hemispheres on each lattice site is shown in **Figure 3d**. The periodic coupling of the traps results in the formation of a band-structure, with a curvature dependent on the coupling between the lenses. The bands still show some discretization due to the finite size of our chain. At the same time, it shows a distinct mode hybridization into polaritonic bands, which can be achieved when using this DBR-based approach. Another work using a perylene-based dye as emitter material for polaritons in photonic potentials created by laser patterning of a surrounding polystyrene layer [67] could show discretized modes and a band structure by expanding the resonators to a 2D square lattice. In our case, an s-band and a p-band, as well as a distinct bandgap between the two bands can be observed in the dispersion of **Figure 3d**. The shape of the band-structure closely resembles the textbook tight-binding model for a finite lattice with a one-atomic basis, as shown with a fit of the s-band marked with a dotted line. From the fit, we estimate a nearest neighbor hopping strength of 2.0 meV which is of the same order as our earlier works [62,69]. The integrated real-space image of the binding s-band is shown in **Figure 3e** and clearly shows the hybridized s-orbitals as introduced in the inset of **Figure 3b**.

Our results pave the way for demonstrating exciton-polariton condensation inside one-dimensional chains as shown by Dusel *et al.* [68-69] or e.g. in perovskite materials [70] or extending the coupling to complex two-dimensional lattices [71] and study e.g. topological effects in lattices [72-74] with PBI containing resonator structures in future. Furthermore, the coupling between the polariton pseudo-spin and its orbital angular momentum, which is accessible via TE-TM splitting of the cavity mode and the corresponding polarization of the decaying polaritons can be investigated. Interestingly, in the context of lattice emulation, circular polarization can serve as a (pseudo)-spin and adds another important degree of freedom to polariton systems. This gives rise to a variety of effects such as Spin-Hall effect in Rashba-Dresselhaus regime [75] or non-Hermitian skin effect [76] that can readily be studied with PBI-based polariton lattices.

For these applications, the great advantage of PBIs next to their long-term stability is the precise control of molecular packing in the solid state by chemical functionalization of the chromophore π-core. This allows adjustment of their molecular optical properties such as the absorption and PL maxima, their respective linewidths, as well as PLQY and excited state lifetime ($\tau_{PL}$). Additional tunability of these photophysical properties can be realized by precise design of the molecular interactions within defined molecular aggregates [77] or (liquid-)crystalline solid states of these emissive materials [45, 78-80] as well as hybrid materials [81], giving rise to exciton diffusion length up to 180 nm [78] or anisotropic layer orientation with unique photoconductivity [33, 82]. Our work clearly demonstrates the high level of technological control over microcavity fabrication based on PBIs as well as the large variety in terms of modulating the photonic potential. For future works, we want to highlight the potential of controlling the excitonic counterpart of exciton polaritons independently by chemical functionalization.

## 3. Conclusion

In this work, we have introduced 1,7-disubstituted perylene bisimide (**PBI-1**) as a highly stable organic solid-state emitter material for strong light-matter coupling at ambient conditions in optical microcavities. We unequivocally demonstrate polariton lasing using the enwrapped PBI derivative with high photoluminescence quantum yield. Furthermore, we demonstrate a high level of technological control over the layer deposition as well as the cavity design and implement individual resonator traps as well as an optical resonator lattice yielding a distinct polaritonic band structure. While conventional crystalline semiconductor materials are limited by lattice matching constraints, the dielectric cavities hosting complex organic emitter materials presented in this work provide a highly flexible platform for fundamental studies of light-matter interaction as well as new optoelectronic device concepts. Our results pave the way for future research on PBI CT-complexes and J-aggregates in aligned thin films as well as chiral PBI dyes for circular polarized luminescence in various cavity approaches such as e.g. open cavities. Finally, the possibility of realizing extended photonic lattices gives rise to exciting applications such as topological lasers.

**Material and Methods**

Sample Preparation: The glass substrates were cleaned in water, acetone and isopropyl alcohol for 10 minutes each and subsequent oxygen plasma treatment for 2 minutes. After the cleaning process, the photonic structures such as single hemispheres and lattices were milled into the top substrate with Gallium ions of a FEI Helios Dual Beam system using a nominal emission current of 6.5 nA, an acceleration voltage of 30 kV, a dwell time of 15 μs and 20 passes. The dielectric mirrors were deposited on the prepared bottom and top glass substrates by dual ion beam sputtering (Nordiko 3000), utilizing an assist ion source. In this case, 9 bilayers of $SiO_2$/$TiO_2$ with a thickness of 93.4 nm and 60 nm respectively were deposited. The last layer of $TiO_2$ was capped with 20 nm of $SiO_2$.

For the synthesis of **PBI-1** solvents and reagents were obtained commercially and used as received without further purification unless otherwise stated. The synthesis of *N,N'*-dicyclohexyl-1,7-bis(2,4,6-tris-(3,5-di-*tert*-butylphenyl)-phenoxy)perylene-3,4:9,10-bis(dicarboximide) as well as its singly crystal structure (CCDC1901807) along with photophysical investigation were reported previously [47]. The Spin coating (POLOS™, SPS-Europe spin coater) was performed from solutions of **PBI-1** (4 mg mL$^{-1}$) in dichloromethane (anhydrous grade, Sigma Aldrich). A solution volume of 40 μL was applied to the bottom DBR using a static dispense method at 1500 rpm s$^{-1}$ and 1500 rpm for 30 s. Subsequentially, the thin films were annealed at 150 °C for 5 min on a Harry Gestigkeit GmbH precision heating stage. To assemble the cavity, 3 μL of BSA (bovine serum albinum, dissolved in water, 200 mg/ml) were drop casted onto the bottom mirror with the spincoated layer of 25 nm **PBI-1** on top. The second, planar or structured mirror (depending on the sample) was placed on the BSA film, pressed in place and left to dry under a force of about 0.25 N cm$^{-2}$ for 2 days.

Transmission electron microscope (TEM) and Selected area electron diffraction (SAED): SAED measurements were conducted with FEI Titan 80-300 TEM operated with an accelerating voltage of 300 kV. The thin films of **PBI-1** for the SAED experiments were prepared by spin coating as described above from $CHCl_3$ (anhydrous grade, Sigma Aldrich®) solution onto PEDOT:PSS coated glass substrates and successive thermal annealing. Afterwards, the substrates were immersed in purified water and the floating PBI film was taken from the water with carbon-covered copper grids for TEM (Lacey Carbon Films on 200 Mesh Copper Grids, Agar Scientific Ltd.).

Experimental Setup:

In the experimental setup, a supercontinuum white-light source filtered with RF to a wavelength of 480 nm for sample excitation was used (NKT SuperK Fianum). The excitation beam was directed onto the sample surface using a high numerical aperture (NA = 0.65) objective (50x) resulting in a laser spot size of d~3 μm in diameter. The emission from the sample was collected in reflection geometry through the same objective, filtered with a 500 nm long pass filter, and directed to the entrance slit of a 500 mm Czerny–Turner spectrometer (Andor Shamrock 500i) where the signal was recorded with a Peltier-cooled EMCCD (Andor Newton 971). The spectrometer was equipped with three different gratings (150 lines $mm^{-1}$, 300 lines $mm^{-1}$, and 1200 lines $mm^{-1}$) and a motorized entrance slit, resulting in a spectral resolution of up to 200 μeV for energies around 2 eV. For white-light reflection measurements, an LED lamp (Thorlabs MCWHL-8) is used. Angle-resolved measurements are carried out in a Fourier imaging configuration using an additional 25 cm focal length lens.

For mode tomographies, the imaging lens is positioned in front of the spectrometer and mounted on a motorized linear translation stage. This allows the collection of slices in real or momentum space at a fixed x or $k_x$ value. These slices are then combined into a 4D matrix, representing the intensity as a function of (x, y) and energy. This approach enables the generation of energy-resolved images in real space without being limited to a one-dimensional cross-section.

**Acknowledgements**

D.H., J.D., S.B., S.H., and S.K. acknowledge financial support by the German Research Foundation (DFG) under Germany's Excellence Strategy–EXC2147 "ct.qmat" (project id 390858490). T.S., M.S., and F.W. thank the Bavarian Ministry for Science and the Arts for the research program "Solar Technologies Go Hybrid".

**Data availability statement**

The data that supports the findings of this study are available from the corresponding author upon reasonable request.

**Supporting Information for "Enwrapped Perylene Bisimide Enables Room Temperature Polariton Lasing and Photonic Lattices"**

*Dominik Horneber\*, Johannes Düreth\*, Tim Schembri, Simon Betzold, Matthias Stolte, Sven Höfling, Frank Würthner, and Sebastian Klembt*

D. Horneber, J. Düreth, Dr. S. Betzold, Prof. Dr. S. Höfling, Prof. Dr. S. Klembt
Technische Physik, Wilhelm-Conrad-Röntgen-Research Center for Complex Material Systems, and Würzburg-Dresden Cluster of Excellence ct.qmat, University of Würzburg, 97074 Würzburg, Germany
\*These authors have contributed equally.
E-mail: sebastian.klembt@uni-wuerzburg.de

T. Schembri, Dr. M. Stolte, Prof. Dr. F. Würthner
Institut für Organische Chemie and Center for Nanosystems Chemistry (CNC), University of Würzburg, 97074 Würzburg, Germany
E-mail: frank.wuerthner@uni-wuerzburg.de

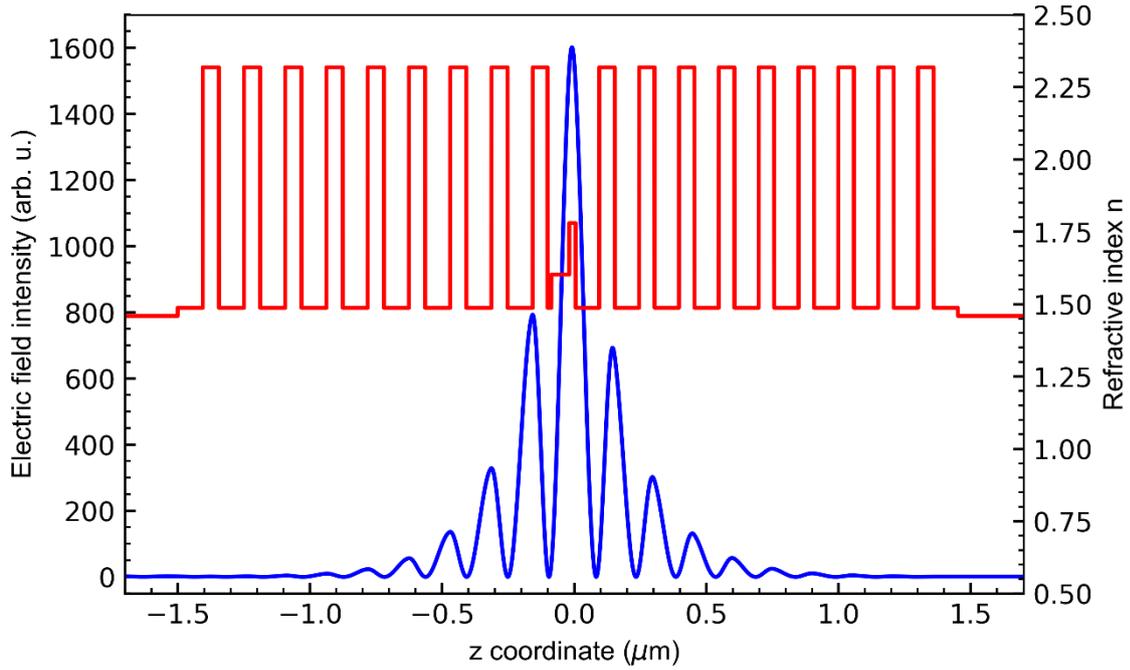

**Figure S1:** Electric field distribution in the microcavity structure as obtained by transfer-matrix method (TMM) calculations highlighting the maximum of the field at the position of our **PBI-1** emitter layer. Refractive indices of the $SiO_2/TiO_2$ DBRs as well as of the BSA and **PBI-1** layer are shown in red.

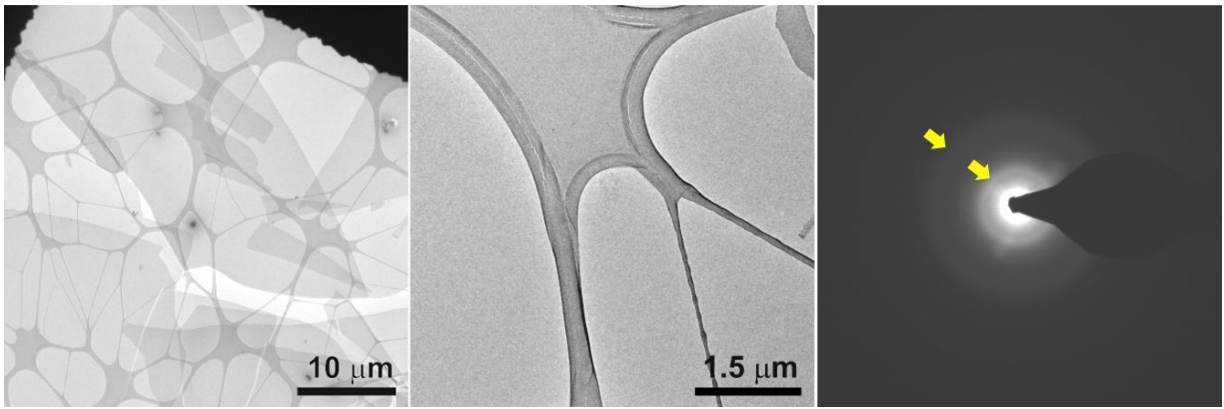

**Figure S2:** TEM images (left, middle) and SAED pattern (right) of a thin film of PBI-1 (25 nm) on Lacey-carbon film. The observed pattern, two wide and diffuse rings (yellows arrows), corresponds to the isotropic nature of the film or the order only over closer distances. The rings are attributed to the following d-spacing values: $d_1 = 4.9$ Å and $d_2 = 2.1$ Å.

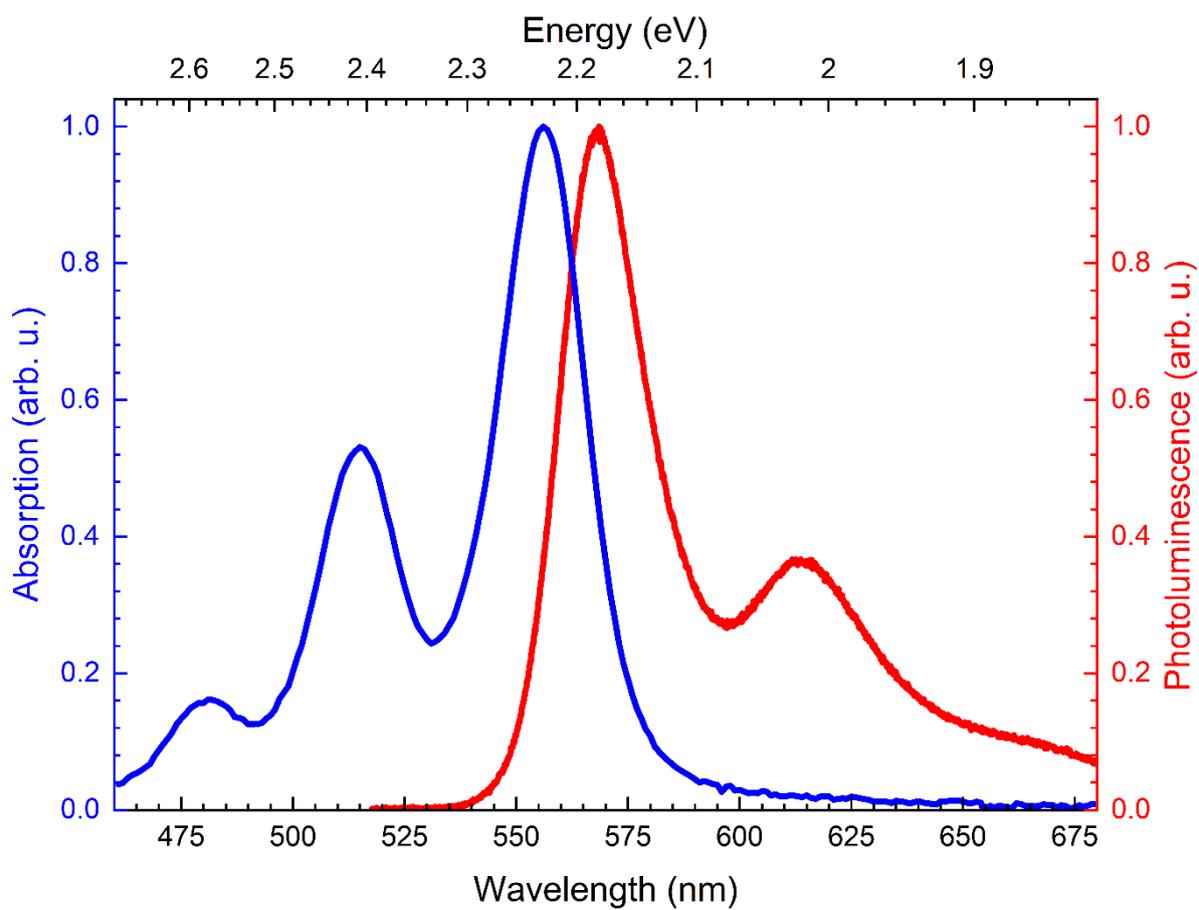

**Figure S3:** Normalized absorbance (blue) and photoluminescence (red) spectra of a 25 nm thin-film of **PBI-1** on GaAs substrate.

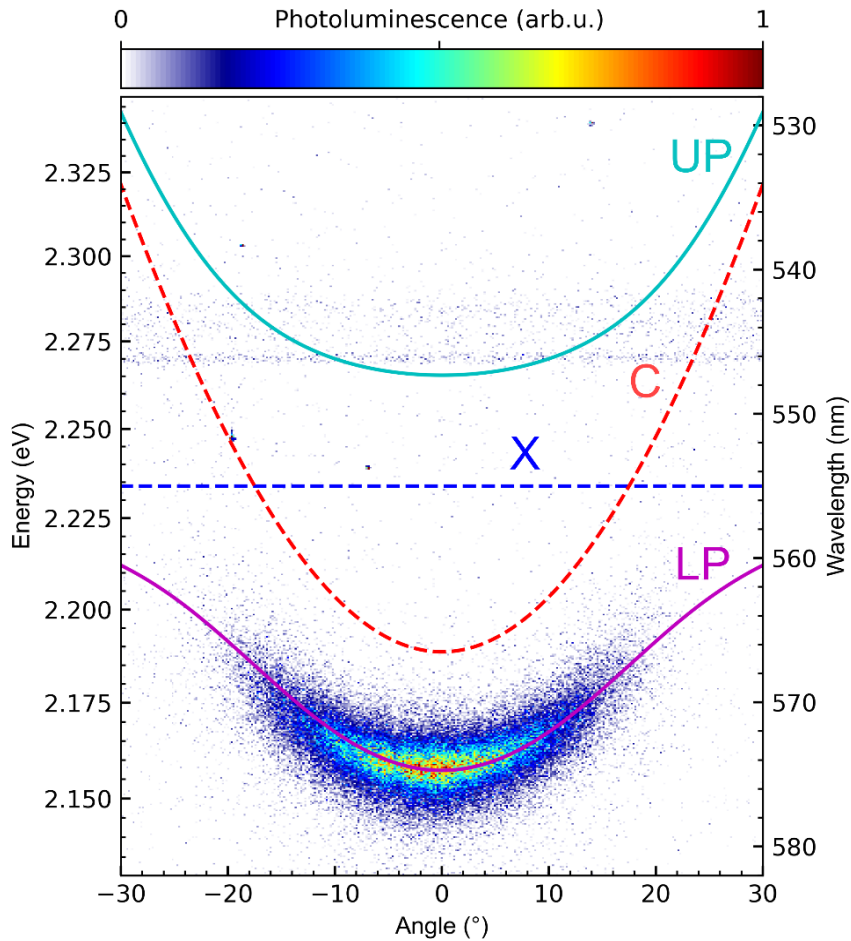

**Figure S4:** Angle-resolved photoluminescence of the stacked planar microcavity with a 25nm solid-sate emitter layer of **PBI-1**. The coupled harmonic oscillator fit reveals a Rabi splitting of $\hbar\Omega_R$ = 98 meV and agrees well with the reflection measurement and simulation data shown in the main text. X and C depict the energetic position of the excitonic and photonic cavity mode, respectively, while UP and LP are the upper and lower polariton branches of the strongly coupled system.

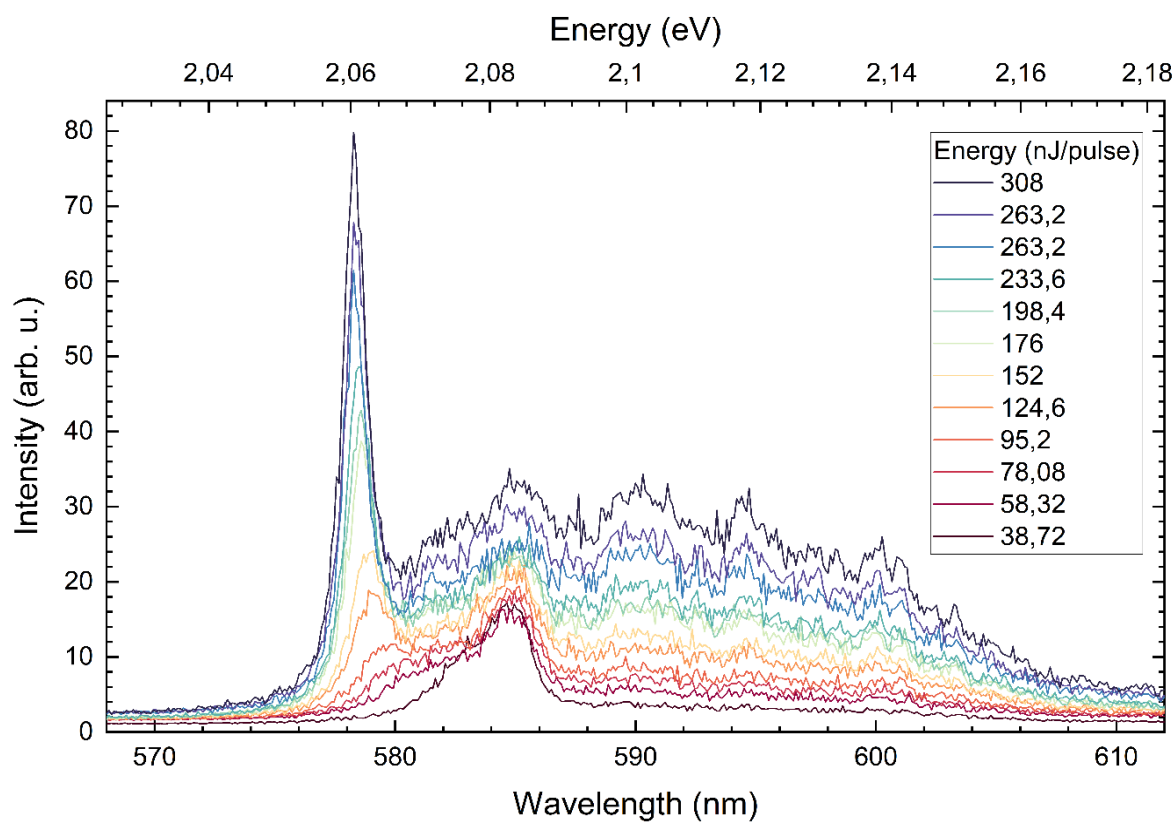

**Figure S5:** Waterfall plot of the PL spectra of the planar stacked **PBI-1** cavity at zero angle for different excitation energies. For low energies starting from 38.72 nJ/pulse (purple) one can see the lower polariton branch around 585 nm which shifts continuously hypsochromically for higher powers up to 308 nJ/pulse (black) and shows a non-linear increase in intensity.